\documentclass[a4paper,11pt]{article}
\usepackage{pos}

\title{Revisiting the $\Lambda_c^+\to \Lambda \eta \pi^+$ and the roles of  intermediate resonances}

\author*[a]{En Wang}
\author[a]{Wen-Tao Lyu}
\author[b]{Man-Yu Duan}
\author[c,d]{Chu-Wen Xiao}
\author[b]{Dian-Yong Chen}
\author[e,f,g]{Ju-Jun Xie}
\author[c,h]{Eulogio Oset}

\affiliation[a]{School of Physics, Zhengzhou University, Zhengzhou 450001, China}

\affiliation[b]{School of Physics, Southeast University, Nanjing 210094, China}

\affiliation[c]{Department of physics, Guangxi Normal University, Guilin 541004, China}

\affiliation[d]{Guangxi Key Laboratory of Nuclear Physics and Nuclear Technology, Guangxi Normal University, Guilin 541004, China}

\affiliation[e]{Southern Center for Nuclear-Science Theory (SCNT), Institute of Modern Physics, Chinese Academy of Sciences, Huizhou 516000, China}
    \affiliation[f]{Institute of Modern Physics, Chinese Academy of
	Sciences, Lanzhou 730000, China} \affiliation[g]{School of Nuclear Sciences and Technology, University of Chinese Academy of Sciences, Beijing 101408, China}

 \affiliation[h]{Departamento de Física Teórica and IFIC, Centro Mixto Universidad de Valencia-CSIC Institutos de Investigación de Paterna, 46071 Valencia, Spain}
 
\emailAdd{wangen@zzu.edu.cn}
\emailAdd{xiaochw@gxnu.edu.cn}
\emailAdd{chendy@seu.edu.cn}
\emailAdd{xiejujun@impcas.ac.cn}
\emailAdd{oset@ific.uv.es}

\abstract{In this paper, we first review the theoretical and experimental studies of the process $\Lambda_c^+ \to \pi^+ \eta \Lambda$. Motivated by the recent BESIII and Belle measurements, we have  conducted a theoretical study of the process $\Lambda_c^+ \to \pi^+ \eta \Lambda$, where the $a_0(980)$ and $\Lambda(1670)$ resonances are dynamically generated from the $S$-wave meson meson and meson baryon interaction, respectively.  Our results give a reasonable description of the invariant mass distributions, which implies that the spin flip contribution in the $\Sigma(1385)$ excitation plays an important role.}

\FullConference{The 21st International Conference on Hadron Spectroscopy and Structure (HADRON2025)\\
 27 - 31 March, 2025\\
Osaka University, Japan\\}

\begin{document}
\maketitle

\section{Introduction}
Hadronic decays of charmed hadrons provide a unique window into hadron-hadron interactions and CP violation phenomena~\cite{Cheng:2015iom}, and serve as critical testing grounds for quantum chromodynamics (QCD) and short-range weak interactions~\cite{Miyahara:2015cja,Duan:2024okk,Li:2024rqb,Lyu:2024qgc,Liu:2023jwo,Zeng:2020och,Wang:2020pem,Zhang:2024jby}. 
The process $\Lambda_c^+\to \Lambda\eta\pi^+$ has called many attentions, since it could provide an important platform to study the light hadrons.

In Ref.~\cite{Xie:2016evi}, the authors have revisited the process $\Lambda_c^+\to \Lambda\eta\pi^+$ taking into account the $\eta \Lambda$ and $\pi^+\eta$ final-state interactions, which give the line shapes of the $\Lambda(1670)$ and  $a_0(980)$ state. Subsequently, Ref.~\cite{Xie:2017xwx} has proposed that this process could be used to search for the $\Sigma(1/2^-)$, which is crucial to understand the the properties of low-lying excited baryons~\cite{Roca:2013cca,Wang:2024jyk,Liu:2017hdx,Wang:2015qta}.

Experimentally, the decays $\Lambda_c^+\to \Lambda\eta\pi^+$ have been studied at the CLEO experiment in 1995~\cite{CLEO:1995cbq} and
2003~\cite{CLEO:2002aec}. 
In 2019, the BESIII has studied the process $\Lambda_c^+\to \Lambda\eta\pi^+$ based on $\Lambda_c^+\bar\Lambda_c^-$ pairs produced  at a center-of-mass energy of $\sqrt{s}=4.6$~GeV with an integrated luminosity of 567~pb$^{-1}$~\cite{BESIII:2018qyg}, and measured the branching fractions  $\mathcal{B}(\Lambda_c^+\to \Lambda\eta\pi^+)=(1.84\pm 0.21(stat)\pm 0.15(syst))\%$. The measured invariant mass distributions show clear peaks of the $\Sigma(1385)$ and $a_0(980)$~\cite{BESIII:2018qyg}.
In 2021, the Belle Collaboration has measured the process $\Lambda_c^+\to \Lambda\eta\pi^+$ based on a 980~fb$^{-1}$ data sample, and determined the absolute branching fraction  $\mathcal{B}(\Lambda_c^+\to \Lambda\eta\pi^+)=(1.84\pm 0.02(stat)\pm 0.09(syst))\%$~\cite{Belle:2020xku}. 
In Ref.~\cite{Yu:2020vlt}, it is found that the final state interaction can give a significant contribution, where $\Sigma^+(1385)$ and $\eta$ in $\Lambda_c^+\to \Sigma^+(1385)\eta$ by exchanging a charged pion are transformed as $\Lambda$ and $a_0(980)^+$, respectively, and the branching fraction is predicted to be $\mathcal{B}(\Lambda_c^+\to \Lambda a_0(980)^+)=(1.7_{-1.0}^{+2.8}\pm0.3)×10^{-3}$. However, it is easy to see that the triangle is very far away from developing a triangle singularity, which indicates that this production mode is relatively inefficient.

In Ref.~\cite{Wang:2022nac}, we have reanalyzed this process by considering the $S$-wave $\eta\Lambda$ and $\eta\pi^+$ final state interactions within the chiral unitary approach, which dynamically generate the $\Lambda(1670)$ and $a_0(980)$, respectively, and our results could well reproduce the Belle measurements of the $\eta\Lambda$ and $\pi^+\Lambda$ invariant mass distributions. Furthermore, we have predicted a clear cusp structure of $a_0(980)$ in the $\pi^+\eta$ invariant mass distribution.
However, the Dalitz plot of the process $\Lambda^+_c\to \eta\pi^+\Lambda$ measured by Belle has more yield appearing in the region of $M^2_{\Lambda\pi^+}<1.85$~GeV$^2$ and $M^2_{\eta\Lambda}<4$~GeV$^2$~\cite{Belle:2020xku}, which cannot be well-described in Ref.~\cite{Wang:2022nac}. The discrepancies may imply some contributions from the resonance $\Sigma^*(1/2^-)$~\cite{Xie:2017xwx}.
In principle, the signal of $\Sigma^*(1/2^-)$ could overlap with the one of $\Sigma(1385)$, and will be difficult to be directly observed in the $\Lambda\pi^+$ invariant spectrum. However, the contributions from $\Sigma^*$ could affect the invariant mass spectrum of $\pi^+ \eta$ correctly. Furthermore, we have considered the predicted low-lying baryon $\Sigma^*(1/2^-)$~\cite{Lyu:2024qgc}, and found that our results involving the $\Sigma^*(1/2^-)$ are favored by fitting to the Belle data of the $\eta\Lambda$ and $\pi^+\Lambda$ invariant mass distributions. Meanwhile, we have  predicted the $\eta\pi^+$ invariant mass distribution and the angular distribution $d\Gamma/d{\rm cos}\theta$, which are significantly different depending on whether or not the contribution from the $\Sigma^*(1/2^-)$ is considered.  Finally, we show that, with the contribution from the $\Sigma^*(1/2^-)$, the calculated Dalitz plot agrees with the Belle measurements~\cite{Lyu:2024qgc}.

The reaction has been posteriorly measured by the BESIII based on 6.1~fb$^{-1}$ of $e^+e^-$ annihilation data collected at center-of-mass energies from 4.600 to 4.843~GeV ~\cite{BESIII:2024mbf} and a wider range of masses in the three invariant mass distributions, $\pi^+\Lambda$, $\pi^+\eta$, $\eta\Lambda$ is presented. In spite of the small signal seen in the Dalitz plot of Belle~\cite{Belle:2020xku}, the analysis of Ref.~\cite{BESIII:2024mbf} reports a branching fraction of 54\% for the $\Lambda a_0^+(980)$ decay mode. This surprising conclusion deserves a detailed attention. While in Ref.~\cite{Xie:2017xwx} the full relativistic propagator of the $\Sigma(1385)$ is taken into account, in Ref.~\cite{Wang:2022nac} only the spin independent part of the $\Sigma^*$ propagator is considered. We shall see that the spin flip part of this amplitude plays an important role. Meanwhile, the branching fraction of $\Lambda_c^+\to\Lambda a_0^+$ is also important to understand the mechanism of the process $\Lambda_c^+\to\pi^+ K^-p$~\cite{Zhang:2024jby}. Thus, we will investigate the process $\Lambda_c^+\to\pi^+ K^-p$ based on the recent BESIII measurements~\cite{BESIII:2024mbf}.

\section{Formalism}
We start from the external emission mechanism of Fig.~1 of Ref.~\cite{Duan:2024czu}  with the hadronization of the strange quark and the $ud$ quark pair, and could obtain all the possible final hadrons,
\begin{equation}\label{Eq:H}
	H=\pi^+\left\{\frac{1}{\sqrt{2}}K^-p+\frac{1}{\sqrt{2}}\bar{K}^0n+\frac{1}{3}\eta\Lambda\right\},
\end{equation}
which has to be considered as the tree level contribution.
We can see that the last term in Eq.~(\ref{Eq:H}) is $\frac{1}{3}\pi^+\eta\Lambda$, which directly corresponds to the desired final state. 
It is also important to note that the $\bar{K}N$ channels could transit to $\eta\Lambda$ and reach the $\pi^+\eta\Lambda$ final state. In addition, the rescattering $\eta\pi^+\to\eta\pi^+$ rescattering should also be considered in the chiral unitary approach for meson meson interaction. The details can be found in Ref.~\cite{Duan:2024czu}.

The mechanism of $\Sigma(1385)$ excitation, as depicted in Fig.~3 of Ref.~\cite{Duan:2024czu}, requires making a transition from a baryon state of $1/2^+$ to another one of $3/2^+$, 
\begin{equation}
	\left\langle\frac{3}{2}M\left|S^\dagger_\nu\right|\frac{1}{2}m\right\rangle=C\left(\frac{1}{2},\mathbf{1},\frac{3}{2}; m,\nu,M\right)\left\langle\frac{3}{2}\left| \left|  S^\dagger \right|\right|\frac{1}{2}\right\rangle,
\end{equation}
with $\nu$ the spherical component of the rank $\mathbf{1}$ operator $\vec{S}^\dagger$. 
The amplitude for $\Sigma(1385)$ excitation, with a $P$-wave coupling in each of the $1/2\to3/2$ transition vertices, is given by
\begin{eqnarray}
	t_{\Sigma(1385)}&=&\alpha \left\langle m^\prime\left|S_i\vec{P}^*_{\pi^+ i}\right|M\right\rangle\left\langle M\left|S_j^\dagger\vec{P}_{\eta j}^*\right|m\right\rangle D  \nonumber \\
	&=&\alpha D\left\langle m^\prime\left|\frac{2}{3}\delta_{ij}-\frac{i}{3}\epsilon_{ijs}\sigma_s\right|m\right\rangle \vec{P}^*_{\pi^+ i}\vec{P}_{\eta j}^* \nonumber \\
	&=&\alpha D\left\langle m^\prime\left|\frac{2}{3}\vec{P}^*_{\pi^+}\cdot\vec{P}_{\eta}^*-\frac{i}{3}\epsilon_{ijs}\sigma_s\vec{P}^*_{\pi^+ i}\vec{P}_{\eta j}^*\right|m\right\rangle,
\end{eqnarray}
where 
\begin{equation}
	D=\dfrac{1}{M_{\text{inv}}(\pi^+\Lambda)-M_{\Sigma(1385)}+{i\Gamma_{\Sigma(1385)}}/{2}}.
\end{equation}

We write the $\Lambda_c^+\to\pi^+\eta\Lambda$ decay amplitude as
\begin{equation}
	t=t_1+t_2,
\end{equation}
with
\begin{eqnarray}\label{Eq:t1}
	t_{1}&=&A \left\{ h_{\pi^+\eta\Lambda}+h_{\pi^+\eta\Lambda}G_{\eta\Lambda}(M_{\text{inv}}(\eta\Lambda))t_{\eta\Lambda,\eta\Lambda}(M_{\text{inv}}(\eta\Lambda))\right. \nonumber\\
	&&+h_{\pi^+\eta\Lambda}G_{\pi^+\eta}(M_{\text{inv}}(\pi^+\eta))t_{\pi^+\eta,\pi^+\eta}(M_{\text{inv}}(\pi^+\eta))  \nonumber\\
	&&+h_{\pi^+\bar{K}N}G_{K^-p}(M_{\text{inv}}(\eta\Lambda))t_{K^-p,\eta\Lambda}(M_{\text{inv}}(\eta\Lambda))   \nonumber\\
	&&+h_{\pi^+\bar{K}N}G_{\bar{K}^0n}(M_{\text{inv}}(\eta\Lambda))t_{\bar{K}^0n,\eta\Lambda}(M_{\text{inv}}(\eta\Lambda)) \left.+\frac{\beta}{M_\Lambda}\frac{2}{3}\vec{P}^*_{\pi}\cdot\vec{P}_{\eta}^* D\right\},
\end{eqnarray}
where
\begin{equation}
	h_{\pi^+\eta\Lambda}=\frac{1}{3};~~~h_{\pi^+\bar{K}N}=\frac{1}{\sqrt{2}},~~~	t_2=-\frac{i}{3}\frac{A\beta}{M_\Lambda}\epsilon_{ijs}\sigma_s\vec{P}^*_{\pi i}\vec{P}_{\eta j}^* D,
\end{equation}
and $A$ is a global normalization constant and $\beta$ is a free parameter. $G_i$ and $t_{i,j}$ are the loop function and transition amplitude of the meson-baryon or meson-meson~\cite{Duan:2024czu}. 
According to the notation 1 for $\pi^+$, 2 for $\eta$ and 3 for $\Lambda$ and applying the standard formula of the Review of Particle Physics (RPP), we have
\begin{equation}
	\dfrac{d^2\Gamma}{dM_{12}dM_{23}} = \frac{1}{(2\pi)^3}\dfrac{2M_\Lambda 2M_{\Lambda_c^+}}{32M_{\Lambda_c^+}^3}\overline{\sum}\sum|t|^2~2M_{12}~2M_{23}.
\end{equation}

\section{Results and discussion}

\begin{figure}	
	\centering
		\includegraphics[scale=0.39]{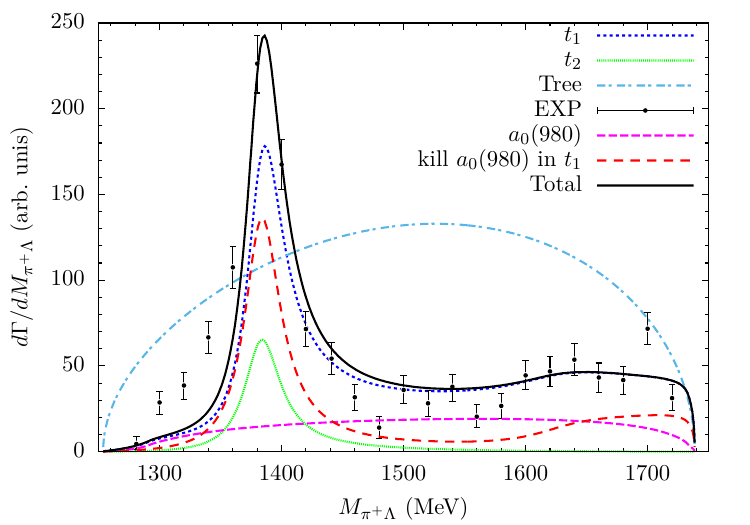}	
		\includegraphics[scale=0.39]{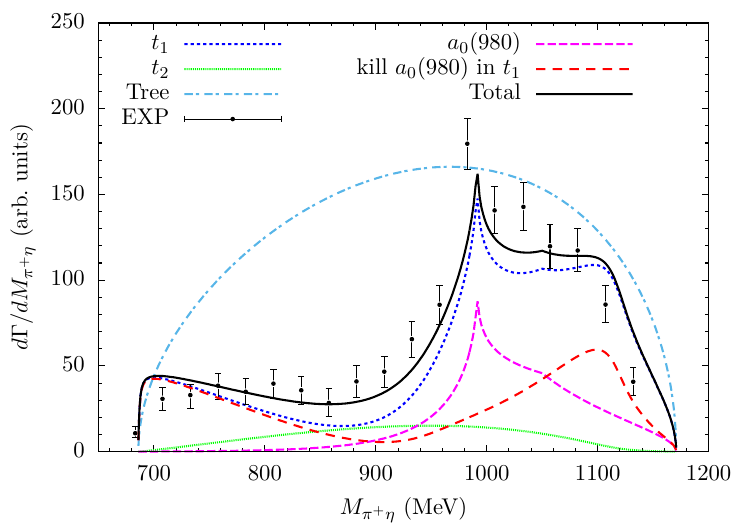}	
		\includegraphics[scale=0.39]{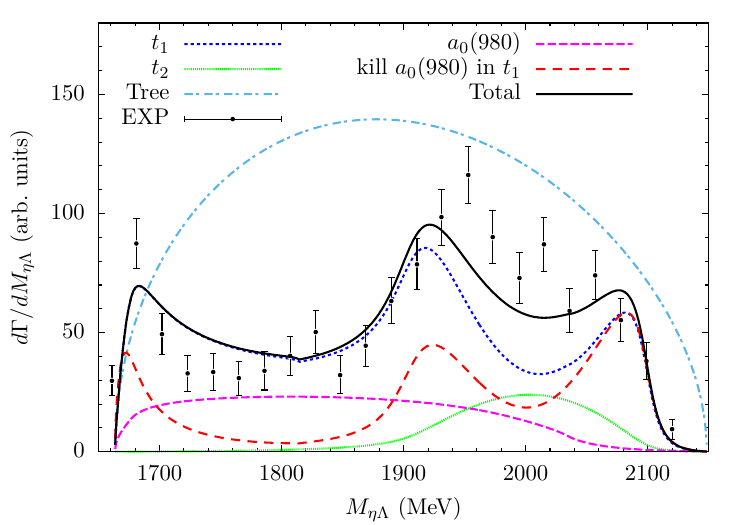}	
	\caption{$\pi^+\Lambda$~(a), $\pi^+\eta$~(b), and $\eta\Lambda$~(c) invariant mass distributions of the $\Lambda^+_c\to \eta\pi^+\Lambda$ decay.}\label{fig:mass-distribuion-without-phase}
\end{figure}
\begin{figure}
	\centering
		\includegraphics[scale=0.39]{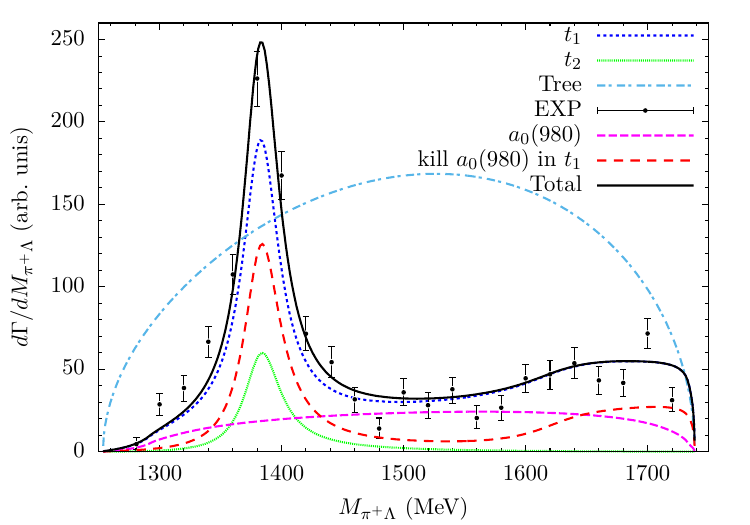}
		\includegraphics[scale=0.39]{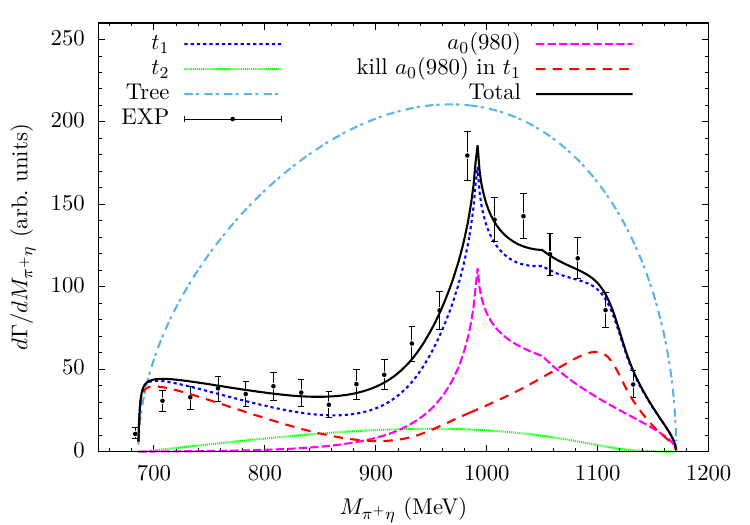}
		\includegraphics[scale=0.39]{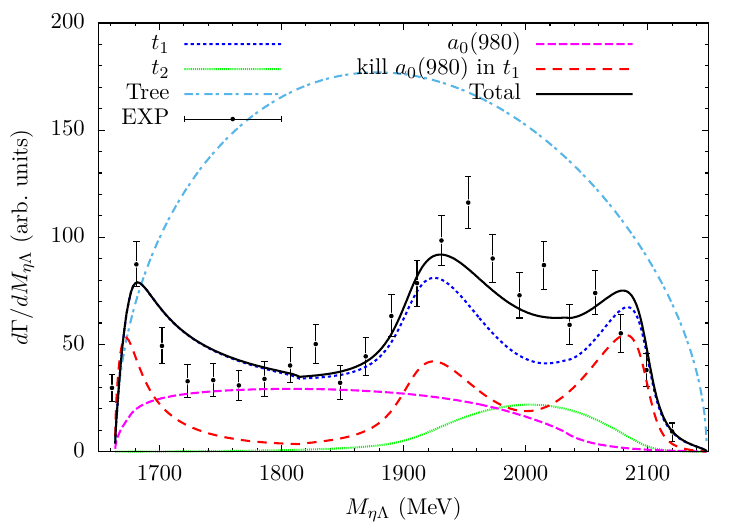}
	\caption{$\pi^+\Lambda$~(a), $\pi^+\eta$~(b), and $\eta\Lambda$~(c) invariant mass distributions of the $\Lambda^+_c\to \eta\pi^+\Lambda$ decay with $e^{i\phi}\beta$.}\label{fig:mass-distribuion-with-phase}
\end{figure}

In Fig.~\ref{fig:mass-distribuion-without-phase} we plot the mass distributions of $\pi^+\Lambda$, $\pi^+\eta$, and $\eta\Lambda$. The parameters of the fit are $A=0.457$, $\beta=0.098$, and $\chi^2/d.o.f.=208.47/ (63-2)=3.42$.

We observe at first glance that the agreement with the three experimental mass distributions is fair, and the basic features are reproduced. In Fig.~\ref{fig:mass-distribuion-without-phase}(a) we see the $\pi^+\Lambda$ mass distribution, which is dominated by the $\Sigma(1385)$ excitation, and we see that both the spin non flip part (present in $t_1$) and the spin flip part present in $t_2$ contribute and produce the same shape in the $\pi^+\Lambda$ distribution. We also observe that the $a_0^+(980)$ contribution coming in our case from the $t_{\eta\pi^+,\eta\pi^+}$ amplitude, has a shape similar to that obtained in the analysis of Ref.~\cite{BESIII:2024mbf} but with smaller strength. 

Next we look at Fig.~\ref{fig:mass-distribuion-without-phase}(b) with the $\pi^+\eta$ mass distribution. We see that we get a signal for $\Lambda a_0^+(980)$ production coming from the $t_{\eta\pi^+,\eta\pi^+}$ amplitude, which, however, has an integrated strength less than one half the one obtained in the analysis of Ref.~\cite{BESIII:2024mbf}. 

Finally, in Fig.~\ref{fig:mass-distribuion-without-phase}(c) we plot the result for the $\eta\Lambda$ mass distribution. We observe that the peak of $\Lambda(1670)$ is automatically generated, while in the BESIII analysis it is a fitting term. The strength of the $\Lambda(1670)$ and $a_0^+(980)$ production are tied in our approach to the strength of the tree level and the $\pi^+K^-p$, $\pi^+\bar{K}^0n$ relative weight to the tree level provided by the weak production mechanism of Eq.~(\ref{Eq:H}). The tree level has a large strength here and very interesting, the spin flip part of the $t_2$ amplitude produces a bump around 2000~MeV.

While the fit obtained in terms of just one parameter, up to the global normalization, reproduces fairly well the features of all the mass distributions, we have taken advantage of the freedom in the phase of the $\beta$ term in Eq.~(\ref{Eq:t1}) and allow $\beta\rightarrow e^{i\phi}\beta$ to acquire a phase.
In this case we have two free parameters and the fit returns the value $A=0.515$, $\beta=0.084$, $\phi=0.44\pi$ and $\chi^2/d.o.f.=125.81/ (63-3)=2.09$. 
We plot the results in Fig.~\ref{fig:mass-distribuion-with-phase}. We see an improvement in the mass distributions and the $\chi^2$ value but the basic features were already obtained in the one parameter fit.

\section{ Conclusions }
In this paper, we have revisited the $\Lambda_c^+ \to \pi^+ \eta \Lambda$ decay observed by the Belle and BESIII Collaborations and conducted a theoretical study of the reaction to pin down the essential elements. In our approach the $a_0(980)$ and $\Lambda(1670)$ resonances are dynamically generated from the $S$-wave meson meson and meson baryon interaction, respectively. 

We have shown that the consideration of the $a_0(980)$ and $\Lambda(1670)$ resonances as dynamically generated from the meson-meson and meson-baryon interactions, respectively, has allowed us to find a reasonable description of the invariant mass distributions for the $\Lambda_c^+ \to \pi^+ \eta \Lambda$ decay in terms of just one parameter, which is associated with the strength of the $\Sigma(1385)$ resonance, that appears in $P$-wave and is not linked to the other $S$-wave amplitudes. We also call the attention to the role played by the spin flip contribution in the $\Sigma(1385)$ excitation.

\section*{Acknowledgments}
This work is supported by the National Key Research and Development Program of China (No. 2024YFE0105200), the Natural Science Foundation of Henan under Grant No. 232300421140, the National Natural Science Foundation of China under Grant No. 12475086, No. 12192263, No. 12175037, No. 12365019, and No. 12335001. This work is also supported by the Spanish Ministerio de Ciencia e Innovaci\'on (MICINN) under contracts PID2020-112777GB-I00, PID2023-147458NB-C21 and CEX2023-001292-S; by Generalitat Valenciana under contracts PROMETEO/2020/023 and CIPROM/2023/59. The work is partly by the Natural Science Foundation of Changsha under Grant No. kq2208257, the Natural Science Foundation of Hunan Province under Grant No. 2023JJ30647, the Natural Science Foundation of Guangxi Province under Grant No. 2023JJA110076.

\end{document}